\documentstyle[12pt]{article}

\begin{document}


\begin{centering}
{\leftskip=2in \rightskip=2in
{\large \bf Fundamental physics in space:}}\\
{\leftskip=2in \rightskip=2in
{\large \bf a Quantum-Gravity perspective\footnote{To appear in a special
issue ``Fundamental physics on the International Space Station and in space"
of {\it General Relativity and Gravitation}.}}}\\
\bigskip
\bigskip
\medskip
{\bf Giovanni AMELINO-CAMELIA}\\
\bigskip
{\it Dipart.~Fisica Univ.~La Sapienza and Sez.~Roma1 INFN}\\
{\it P.le Moro 2, I-00185 Roma, Italy}

\end{centering}

\vspace{0.7cm}
\begin{center}
{\bf ABSTRACT}
\end{center}

\baselineskip 11pt plus .5pt minus .5pt

{\leftskip=0.6in \rightskip=0.6in
{\footnotesize I consider the possibility
that space experiments be used to search for quantum properties
of spacetime. On the basis of recent quantum-gravity results,
I argue that insight on some quantum properties of spacetime
can be obtained with experiments planned for
the International Space Station,
such as AMS and EUSO, and with satellite gamma-ray telescopes,
such as GLAST.
}}

\newpage
\baselineskip 12pt plus .5pt minus .5pt
\pagenumbering{arabic}

\setcounter{footnote}{0}
\renewcommand{\thefootnote}{\alph{footnote}}

\pagestyle{plain}

\section{Introduction}
Any attempt to test new ideas about the
fundamental laws of Nature requires remarkable
accuracy and/or the study of particles with very
high energies.
Since these objectives are often beyond
the reach of conventional laboratories,
there has been growing interest in
the analysis of the implications of fundamental-physics theories
in astrophysics and cosmology. More and more examples are being found
in which a given picture of the fundamental laws of physics is established
to have important consequences for certain classes of observations
in astrophysics and cosmology. At first these studies
focused on the implications of particle-physics theories, and
the field of ``astroparticle physics" is now
one of the leading areas of fundamental-physics research.
Over the last few years
it has been realized that even the study
of certain quantum-gravity
models (which typically pertain to energy scales even beyond the
ones of interest in particle physics) can benefit from
suitable analyses of astrophysics and cosmology contexts.

Space experiments play an important role in these studies.
Especially for certain observations in astrophysics, it is important
to be able to observe the process before the disturbance introduced
by the atmosphere. Moreover, in addition to its importance in astrophysics
and cosmology, the opportunity to conduct experiments
in space is also sometimes exploited in setting up high-precision
experiments. The quieter environment of a space laboratory
is in fact very valuable for achieving high precision in certain
experiments.

Most of the experimental
studies that are being considered in ``quantum-gravity
phenomenology"~\cite{polonpap}
rely on space experiments.
In quantum-gravity phenomenology
one is looking for the small effects predicted by quantum-gravity
theories, effects with magnitude set by the
ratio between the energy of the particles involved and
the huge Planck energy scale ($E_p \simeq 10^{28}eV$).
The overlap of interests between astrophysics and
quantum-gravity phenomenology comes primarily
from the availability in astrophysics of particles with
energies significantly higher than the energies that
can be achieved at particle colliders.
And in quantum-gravity phenomenology there is of course
strong interest in high-precision experiments, since the
Planck-scale effects are always very small.

I here want to discuss some of the key aspects of this
quantum-gravity phenomenology in space, with emphasis on
some studies which can be conducted on the ``ISS",
the International Space Station.
In the first part of this paper
(Sections~2-7) I review some results
obtained in the quantum-gravity literature
that are relevant for space experiments.
I will stress that a key indication which is emerging from
quantum-gravity research is that it is natural to question
whether in quantum spacetime it is possible for
ordinary (classical) Lorentz symmetry to preserve
the role it holds in current (pre-quantum-gravity) theories.
In light of these results, certain types of Lorentz-symmetry tests
acquire interest from a quantum-gravity perspective.
In addition to discussing the fate of Lorentz symmetry in
quantum spacetime, I will also (more briefly) comment on another
type of effect which has been considered in the quantum-gravity
literature, the possibility that in monitoring a given length
one might encounter an irreducible level of length fluctuations
(connected with a sort of uncertainty principle
for length measurements).
In the second part of the paper (Section~8)
I will discuss some characteristic
signature of these quantum-gravity effects and comment
on some space experiments that represent
opportunities for searches of these quantum-gravity effects.

\section{Lorentz symmetry and the
three perspectives on the Quantum Gravity problem}
It is probably fair to state that each quantum-gravity research line
can be connected with one of
three perspectives on the problem: the particle-physics perspective,
the general-relativity perspective and the condensed-matter perspective.

From a particle-physics perspective it is natural to attempt to
reproduce as much as possible the successes of the Standard Model
of particle physics.
One is tempted to see gravity simply as one more gauge interaction.
From this particle-physics perspective a
natural solution of the quantum-gravity problem should have
 its core features
described in terms of graviton-like exchange
in a background classical spacetime.
Indeed this structure is found in String Theory, the most developed
among the quantum-gravity approaches that originate from
a particle-physics perspective.

The general-relativity perspective
naturally leads to reject the use of
a background spacetime~\cite{carloreview,leereview}.
According to General Relativity the evolution of particles
and the structure of spacetime are selfconsistently connected:
rather than specify a spacetime arena (a spacetime background) beforehand,
the dynamical equations determine at once both the spacetime structure
and the evolution of particles.
Although less publicized, there is also growing awareness
of the fact that, in addition to the concept of background independence,
the development of general relativity relied heavily
on the careful consideration of the in-principle limitations that
measurement procedures can encounter\footnote{Think for example of the
limitations that the speed-of-light limit imposes on certain setups
for clock synchronization and of the contexts in which it is
impossible to distinguish between a constant acceleration and
the presence of a gravitational field.}.
In light of the various arguments
suggesting that, whenever both quantum mechanics and general relativity
are taken into account, there should be an in-principle Planck-scale
limitation to the localization of a spacetime
point (an event),
the general-relativity perspective
invites one to renounce to any direct reference
to a classical spacetime~\cite{dopl1994,ahlu1994,ng1994,gacmpla,garay}.
Indeed this requirement that spacetime be described as fundamentally
nonclassical (``fundamentally quantum''),
the requirement that the in-principle measurability limitations
be reflected by the adoption of a corresponding
measurability-limited description of spacetime,
is another element of intuition which is guiding quantum-gravity
research from the general-relativity perspective.
This naturally leads one to consider
discretized spacetimes, as in the Loop
Quantum Gravity approach, or noncommutative spacetimes.

The third possibility is a condensed-matter perspective
on the quantum-gravity problem
(see, {\it e.g.}, Refs.~\cite{volovik,laugh}),
in which spacetime itself is
naturally seen as an emerging critical-point entity.
Condensed-matter theorists are used to describe the degrees
of freedom that are measured in the laboratory
as collective excitations within
a theoretical framework whose primary description is given
in terms of much different, and often practically unaccessible,
fundamental degrees of freedom.
Close to a critical point some symmetries arise for the
collective-excitations theory, which do not carry the
significance of fundamental symmetries,
and are in fact lost as soon as the theory is probed somewhat
away from the critical point. Notably,
some familiar systems are known to exhibit special-relativistic invariance
in certain limits, even though, at a more fundamental level,
they are described in terms of a nonrelativistic theory.

Clearly for the condensed-matter perspective
on the quantum-gravity problem
it is natural to see the familiar
classical continuous Lorentz symmetry only as an approximate symmetry.

Results obtained over the last few years (which are partly reviewed
later in these notes) allow us to formulate a similar expectation
from the general-relativity perspective.
Loop quantum gravity and other discretized-spacetime quantum-gravity
approaches appear to require a description of the familiar
(continuous) Lorentz symmetry
as an approximate symmetry, with departures governed by the Planck scale.
And in the study of noncommutative spacetimes some Planck-scale
departures from Lorentz symmetry appear to be inevitable.

From the particle-physics perspective there is instead no obvious reason
to renounce to exact Lorentz symmetry, since
Minkowski classical spacetime is an admissible background spacetime,
and in classical Minkowski there cannot be any a priori obstruction
for classical Lorentz symmetry.
Still, a break up of Lorentz symmetry,
in the sense of spontaneous symmetry breaking,
is of course possible, and this
possibility has been studied extensively
over the last few years, especially in String Theory
(see, {\it e.g.},
Ref.~\cite{dougnekr} and references therein).

\section{Quantum Gravity Phenomenology}
The key challenge for the search of experimental hints relevant
for the quantum-gravity problem is the smallness of the
effects that one might expect to be induced by a quantum gravity.
A key point for this ``Quantum Gravity Phenomenology"~\cite{polonpap}
is that we are familiar with ways to gain sensitivity
to very small effects. For example,
our understanding of brownian motion is based on the fact that
the collective result of a large number of tiny microscopic effects
eventually leads to an observably large macroscopic effect.
The prediction of proton decay within certain grandunified theories
(theories providing a unified description of electroweak and strong
particle-physics interactions) is really a small effect, suppressed
by the fourth power of the ratio between the mass of the proton and the
grandunification scale, which is only
three orders of magnitude smaller than the Planck scale.
In spite of this horrifying suppression,
of order $[m_{proton}/E_{gut}]^4 \sim 10^{-64}$,
with a simple idea we have managed to acquire a remarkable sensitivity
to the possible new effect: the proton lifetime predicted by grandunified
theories is of order $10^{39}s$ and ``quite a few" generations
of physicists should invest their entire lifetimes staring at a single
proton before its decay, but by managing to keep under observation
a large number of protons
our sensitivity to proton decay is significantly increased.
In proton-decay searches the number of protons being monitored
is the (ordinary-physics) dimensionless quantity that works
as ``amplifier" of the new-physics effect.

In close analogy with these familiar strategies for the study of
very small effects, in quantum-gravity phenomenology one
must focus~\cite{polonpap}
on experiments which have something to do with spacetime
structure and that host an ordinary-physics dimensionless quantity
large enough that it could amplify
the extremely small effects we are hoping to discover.
The amplifier will often be the number of small effects contributing
to the observed signal (as in brownian motion and in
proton-stability studies).

Using these general guidelines, a few quantum-gravity
research lines have matured over these past few years.
I will later discuss studies of the fate of Lorentz
symmetry in quantum spacetime, emphasizing the relevance for
observations of gamma rays in astrophysics~\cite{grbgac,billetal}
and the relevance for the analysis of the cosmic-ray
spectrum~\cite{kifu,ita,aus,gactp,jaco,gacpion,orfeupion}.
I will also discuss laser-interferometric tests of Planck-scale
effects, both the ones~\cite{gaclaem} that are
relevant for the study of the fate
of Lorentz symmetry in quantum spacetime and the ones
which explore the possibility of quantum-gravity-induced distance
fluctuations~\cite{gacgwi,ahlunature,nggwi}.

I will not discuss here matter-interferometric limits
on Planck-scale effects and limits
on Planck-scale effects obtained using the sensitivity
to new physics that one
finds naturally in the neutral-kaon system
(see, {\it e.g.},
Refs.~\cite{ehns,huetpesk,kostcpt,emln,peri,floreacpt,peri}).
In fact, for these Planck-scale effects it is not obvious that
space experiments should be preferred to conventional
on-ground laboratory experiments.

I should also stress that I intend to focus here on the possibility
of a ``genuinely quantum" spacetime, a spacetime which will typically
have intrinsic noncommutativity and/or discretization at the Planck
scale.
Interesting ideas about the interplay between gravity and quantum mechanics
which do not require a genuinely quantum spacetime can be found
in Refs.~\cite{anan1,ahluiqgr,gasperiniEP,dharamEP,dharamCOW2,veneziano,lamer}.

\section{A key issue: should we adopt a fundamentally quantum
spacetime?}
As I already stressed, it is rather obvious that from the particle-physics
perspective one would not expect any modification
of Lorentz symmetry, and on the contrary from the condensed-matter
perspective Lorentz symmetry is naturally seen only as an
approximate symmetry.
It is instead less obvious what one should expect for
the fate of Lorentz symmetry in quantum-gravity approaches
based on the general-relativity perspective, and in fact
some key insight, leading to the expectation that departures
from  Lorentz symmetry are usually present, has been gained
only very recently, mostly in the study of loop quantum gravity
and certain noncommutative spacetimes.

A key point that needs to be clarified when approaching the
quantum-gravity problem from the general-relativity perspective
is whether or not one should adopt  a ``genuinely
quantum'' spacetime.
A genuinely quantum spacetime is essentially a spacetime
in which an event (a spacetime point) cannot be sharply localized.
Loop Quantum Gravity (the present understanding of Loop
Quantum Gravity) and certain noncommutative spacetimes
provide examples of genuinely quantum spacetimes.
In the case in which one might be able to introduce coordinates
for the event, in a quantum spacetime it must be impossible
to determine sharply (in the sense of measurement) all of the
coordinates of an event.

String Theory provides, in my opinion,
an example of theory in which the status
of spacetime is still not fully clarified.
One introduces in fact in String Theory
a background spacetime which can be (and usually is)
classical, but then the theory itself tells us that the
position of  a particle cannot be sharply
determined~\cite{venekonmen,dbrscatt1,dbrscatt2,wittenPT}.
This apparent logical inconsistency might suggest that the classical
spacetime background should only be seen as a formal
tool, void of operative meaning, which should be conveniently
replaced by a physically meaningful spacetime picture in
which no classical-spacetime idealization is assumed.

Various arguments suggest~\cite{dopl1994,ahlu1994,ng1994,gacmpla,garay}
that a theory that truly admits
both general relativity and quantum mechanics as appropriate
limits must renounce to any reference to a classical spacetime,
because such a theory is automatically incompatible with
the possibility of localizing sharply a spacetime point.

In Section~6 I will summarize results obtained over the last 3 or 4 years
which suggest that, if spacetime is ``quantum'' in the sense
of noncommutativity or discreteness, the familiar (classical,
continuous) Lorentz symmetry naturally ends up being only an approximate
symmetry of the relevant ``flat spacetime limit'' of quantum gravity.

\section{The three possibilities for the fate
of Lorentz symmetry in quantum gravity}
There are three possibilities for the
fate of Lorentz symmetry in quantum gravity:
unmodified (exact, ordinary) Lorentz symmetry,
broken Lorentz symmetry, and deformed Lorentz symmetry.

It is easy to understand what it means to preserve
Lorentz symmetry without modification.
We are also all familiar with the concept of "broken Lorentz symmetry"
that is being encountered and discussed in some quantum-gravity research
lines. This is completely analogous to the  familiar situation
 in which the presence
of a background selects a preferred class of inertial observers.
For example, there is a modification of
the energy/momentum dispersion relation for light
travelling in water, in certain crystals, and in
other media. Of course, the existence of crystals
is fully compatible with a theoretical framework that is fundamentally
Lorentz invariant, but in presence of the crystal
the Lorentz invariance is manifest only when
different observers take into account the different form that the
tensors characterizing the crystal (or other background/medium)
take in their respective reference systems.
If the observers only take into account the transformation rules
for the energy-momentum of the particles involved in a process
the results are not the ones predicted by Lorentz symmetry.
In particular, the dispersion relation between energy and momentum
of a particle depends on the background (and therefore takes different
form in different frames since the background tensors take different
form in different frames).

While the case in which
Lorentz symmetry is not modified and the case in which Lorentz
symmetry is broken are familiar, the third possibility recently
explored in the quantum-gravity literature, the case of deformed
Lorentz symmetry introduced in Ref.~\cite{gacdsr},
is rather new and it might be useful to describe it here intuitively.
It is the idea that in quantum gravity it might be appropriate
to introduce a second observer-independent scale,
a large-energy/small-length
scale, possibly given by the Planck scale.
It would amount to another step of the same type
of the one that connects Galilei Relativity and
Einstein's Special Relativity: whereas in Galilei Relativity
the (mathematical) description of boost transformations
does not involve any invariant/observer-independent scale,
the observer-independent speed-of-light scale ``$c$'' is encoded in
the Lorentz boost transformations (which can be viewed
as a $c$-deformation of the Galileo boost transformations),
and similarly in the case of a deformed Lorentz symmetry
of the type introduced in Ref.~\cite{gacdsr} there are two scales
encoded in the boost transformations between inertial
observers (observers which are
still indistinguishable, there is no preferred observer).
In addition to the familiar observer-independent velocity scale $c$,
there is a second, length (or inverse-momentum),
observer-independent scale $\lambda$.

In order to provide additional intuition for
the concept of deformed Lorentz symmetry let me consider the
particular case~\cite{gacdsr,dsrnext,leedsr}
in which the deformation involves a new
dispersion relation $m^2 = f(E,p;\lambda)$
with $f(E,p;\lambda) \rightarrow E^2 - p^2$ in the
limit $\lambda \rightarrow 0$.
A modified dispersion relation can also emerge (and commonly
emerges) when Lorentz symmetry is broken, but of course the
role of the modified dispersion relation in the formalism
is very different in the two cases: when Lorentz symmetry is
broken the modified dispersion relation reflects properties
of a background/medium and the laws of boost/rotation transformation
between inertial observers are not modified, while when Lorentz
symmetry is deformed
the modified dispersion relation reflects the properties
of some new laws of boost/rotation transformation between
inertial observers.
This comparison provides an invitation to consider
again the analogy with the transition from Galilei Relativity
to Special Relativity.
In Galilei Relativity, which does not have any
relativistic-invariant scale,
the dispersion relation is written
as $E=p^2/(2m)$ (whose structure fulfills the requirements
of dimensional analysis without the need for dimensionful
coefficients). As experimental evidence in favour of Maxwell equations
started to grow, the fact that those equations involve a
special velocity scale appeared to require (since it was
assumed that the validity
of the Galilei transformations should not be questioned)
the introduction of a preferred class of inertial observers, {\it i.e.}
the ``ether" background.
Special Relativity introduces the first observer-independent scale,
the velocity scale $c$, its dispersion relation
takes the form $E^2 = c^2 p^2 + c^4 m^2$ (in which $c$ plays a crucial
role for what concerns dimensional analysis), and the presence
of $c$ in Maxwell's equations is now understood not as a manifestation
of the existence of a preferred class of inertial observers but rather
as a manifestation of the necessity to deform the Galilei
transformations (the Lorentz transformations are a dimensionful
deformation of the Galilei transformations).
Analogously in some recent quantum-gravity research there has
been some interest in dispersion relations
of the type $c^4 m^2 =E^2 -  c^2 \vec{p}^2 + f(E,\vec{p}^2;E_p)$
and the fact that these dispersion relations involve an absolute energy
scale, $E_p$, has led to the assumption that
a preferred class of inertial observers should be introduced
in the relevant quantum-gravity scenarios.
But, as I stressed in the papers proposing physical theories with
deformed Lorentz symmetry~\cite{gacdsr}, this assumption is not
necessarily correct:
a modified dispersion relation involving two dimensionful scales
might be a manifestation of new laws of transformation between
inertial observers,
rather than a manifestation of Lorentz-symmetry breaking.

\section{Spacetime and Lorentz symmetry in
popular quantum-gravity approaches}

\subsection{Spacetime and Lorentz symmetry in
String Theory}
String Theory is the most mature quantum-gravity approach
from the particle-physics perspective.
As such it of course attempts to
reproduce as much as possible the successes of quantum field theory,
with gravity seen (to a large extent)
simply as one more gauge interaction.
Although the introduction of extended objects (strings, branes, ...)
leads to subtle elements on novelty, in String Theory
the core features of quantum gravity
are essentially described in terms of graviton-like exchange
in a background classical spacetime.

Indeed String Theory does not lead to spacetime quantization,
at least in the sense that its background spacetime
has been so far described as completely classical.
However, this point is not
fully settled: it has been shown that String Theory
eventually leads to the emergence of a fundamental limitation on
the localization of a spacetime
event~\cite{venekonmen,dbrscatt1,dbrscatt2,wittenPT} and this might be
in conflict with the assumption of a physically-meaningful classical
background spacetime.

If eventually there will be a formulation of String Theory
in a background spacetime that is truly quantum, it is likely
that Lorentz symmetry will then not be an exact symmetry
of the theory.
If instead somehow a classical spacetime background can be meaningfully
adopted, of course then there is no {\it a priori} reason to
contemplate departures from Lorentz symmetry:
classical Minkowski spacetime would naturally be an acceptable
background, and a theory in the Minkowski background can be easily
formulated in Lorentz-invariant manner.

Still , it is noteworthy that,
even assuming that it makes sense to consider a classical
background spacetime, the fate of Lorentz symmetry in String Theory
is somewhat uncertain: it has been found that under appropriate
conditions (a vacuum expectation value for certain tensor fields)
Lorentz symmetry is broken in the sense I described above.
In these cases String Theory admits description
(in the effective-theory sense)
in terms of field theory in a noncommutative spacetime~\cite{dougnekr}
with most of the studies focusing on the possibility that
the emerging noncommutative spacetime is ``canonical''
(see Susection~6.3).

In summary in String Theory (as presently formulated, admitting classical
backgrounds) it is natural to expect that Lorentz symmetry be preserved.
In some cases (when certain suitable background/``external'' fields are
introduced) this fundamentally Lorentz-invariant theory can experience
Lorentz-symmetry breaking. There has been so far no significant interest
or results on deformation of Lorentz symmetry in String Theory
(see, however, Ref.~\cite{maggioreSTRINGdsr}).

\subsection{Spacetime and Lorentz symmetry in
Loop Quantum Gravity}
Loop Quantum Gravity is the most mature approach to the quantum-gravity
problem that originates from the general-relativity perspective.
As for String Theory, it must be stressed that the understanding of
this rich formalism is in progress.
As presently understood, Loop Quantum Gravity predicts an
inherently discretized spacetime~\cite{discreteareaLGQ}.
There has been much discussion recently, prompted by the
studies~\cite{gampul,mexweave}, of the possibility that
this discretization might lead to departures from
ordinary Lorentz symmetry.
Although there are
cases in which a discretization is compatible
with the presence of continuous classical
symmetries~\cite{snyder,simonecarlo,areanew},
it is of course natural, when adopting a discretized spacetime,
to put Lorentz symmetry under careful scrutiny.
Arguments presented in Refs.~\cite{gampul,mexweave,thiemLS},
support the idea of broken Lorentz
symmetry in Loop Quantum Gravity.

Moreover, very recently Smolin, Starodubtsev and I proposed~\cite{kodadsr}
(also see the follow-up study in Ref.~\cite{jurekkodadsr})
a mechanism such that Loop Quantum Gravity
would be described at the most fundamental level as a theory that in the
flat-spacetime limit admits deformed Lorentz symmetry.
Our argument originates from the role that certain quantum symmetry groups
have in the Loop-Quantum-Gravity description of spacetime with
a cosmological constant, and observing that in the flat-spacetime limit
(the limit of vanishing cosmological constant)
these quantum groups might not contract to a classical Lie algebra,
but rather contract to a quantum (Hopf) algebra.

In summary in Loop Quantum Gravity the study of the fate of Lorentz
is still at a preliminary stage. All three possibilities
(preserved, broken and deformed) are still
being explored.
It is noteworthy however that until 3 or 4 years ago there was
a nearly general consensus that Loop Quantum Gravity would preserve
Lorentz symmetry, whereas presently the intuition of a majority of experts
has shifted toward the possibility that Lorentz symmetry be broken or
deformed.

\subsection{On the fate of Lorentz symmetry in noncommutative spacetime}
There has been much recent interest in flat noncommutative spacetimes,
as possible quantum versions of Minkowski spacetime.
Most of the work has focused on various parts of the two-tensor
parameter space
\begin{equation}
\left[x_\mu,x_\nu\right] = i {1 \over E_p^2} Q_{\mu \nu}
+ i {1 \over E_p} C^\beta_{\mu \nu} x_\beta ~,
\label{alllp}
\end{equation}
The assumption that the commutators of spacetime coordinates
would depend on the coordinates at most linearly is adopted both for simplicity
and because it captures a very general intuition: assuming that the Planck
scale governs noncommutativity (and therefore noncommutativity should
disappear in the formal $E_p \rightarrow \infty$ limit)
and assuming that the commutators do not involve singular, $1/x^n$, terms
one cannot write anything more general than (\ref{alllp}).

Most authors consider two particular limits~\cite{wessLANGUAGE}:
the ``canonical noncommutative
spacetimes'', with $C^\beta_{\mu \nu} =0$,
\begin{equation}
\left[x_\mu,x_\nu\right] = i \theta_{\mu \nu}
\label{canodef}
\end{equation}
and the ``Lie-algebra noncommutative spacetimes",
with $Q_{\mu \nu} =0$,
\begin{equation}
\left[x_\mu,x_\nu\right] = i \gamma^\beta_{\mu \nu} x_\beta ~
\label{liedef}
\end{equation}
(I am adopting notation
replacing $Q_{\mu \nu}/E_p^2 \rightarrow \theta_{\mu \nu}$
and $C^\beta_{\mu \nu}/E_p  \rightarrow \gamma^\beta_{\mu \nu}$).

An intuitive characterization of the fate of Lorentz symmetry
in canonical noncommutative spacetimes
can be obtained by looking at wave exponentials. The Fourier
theory in canonical noncommutative spacetime is based~\cite{wessLANGUAGE}
on simple wave exponentials $e^{i p^\mu x_\mu}$ and from
the $[x_\mu,x_\nu] = i \theta_{\mu \nu}$
noncommutativity relations one finds that
\begin{equation}
e^{i p^\mu x_\mu} e^{i k^\nu x_\nu}
= e^{-\frac{i}{2} p^\mu
\theta_{\mu \nu} k^\nu} e^{i (p+k)^\mu x_\mu} ~,
\label{expprodcano}
\end{equation}
{\it i.e.} the Fourier parameters $p_\mu$ and $k_\mu$ combine just as
usual, with the only new ingredient of the overall phase factor that depends
on $\theta_{\mu \nu}$.
The fact that momenta combine in the usual way reflects the fact that
the transformation rules for energy-momentum from one
(inertial) observer to another are still the usual, undeformed,
Lorentz transformation rules. However, the product of wave exponentials
depends on $p^\mu \theta_{\mu \nu} k^\nu$, it depends on the ``orientation"
of the energy-momentum vectors $p^\mu$ and $k^\nu$
with respect to the $\theta_{\mu \nu}$ tensor. This is a first indication
that in these canonical noncommutative spacetimes
there is Lorentz symmetry breaking.
The $\theta_{\mu \nu}$ tensor plays the role of a background
that identifies a preferred class of inertial observers.
Different particles are affected by the presence of this background
in different ways, leading to the emergence of different dispersion relations,
as shown by the results~\cite{seibIRUV,susskind,dineIRUV,gacluisa}
of the study of field theories in canonical noncommutative spacetimes.

In canonical noncommutative spacetimes Lorentz symmetry is ``broken''
and there is growing evidence that Lorentz symmetry breaking occurs
for most choices of the tensors $\theta$ and $\gamma$.
It is at this point clear, in light of several recent results,
that the only way to preserve Lorentz symmetry
is the choice $\theta = 0 =\gamma $, {\it i.e.} the case in which
there is no noncommutativity
and one considers the familiar classical
commutative Minkowski spacetime.
Typically Lorentz symmetry is broken, but
recent results suggest that for some special choices of the
tensors $\theta$ and $\gamma$
Lorentz symmetry might be deformed, rather than broken.
In particular, this appears to be the case for the Lie-algebra
$\kappa$-Minkowski~\cite{majrue,kpoinap,gacmaj,lukieFT,gacmich,wesskappa}
noncommutative spacetime ($l,m = 1,2,3$)
\begin{equation}
\left[x_m,t\right] = {i \over \kappa} x_m ~,~~~~\left[x_m, x_l\right] = 0 ~.
\label{kmindef}
\end{equation}

 $\kappa$-Minkowski
is a Lie-algebra spacetime that clearly enjoys classical space-rotation
symmetry; moreover, $\kappa$-Minkowski
is invariant under ``noncommutative translations"~\cite{gacmich}.
Since I am focusing here on Lorentz symmetry,
it is particularly noteworthy that in $\kappa$-Minkowski
boost transformations are necessarily modified~\cite{gacmich}.
A first hint of this comes from the necessity of a deformed
law of composition of momenta, encoded
in the so-called ``coproduct"\cite{majrue,kpoinap}.
One can see this clearly by considering the Fourier tranform.
It turns out~\cite{gacmaj,lukieFT,majoek} that in
the $\kappa$-Minkowski case the correct formulation
of the Fourier theory requires a suitable ordering prescription
for wave exponentials:
\begin{equation}
 :e^{i k^\mu x_\mu}: \equiv e^{i k^m x_m} e^{i k^0 x_0}
~.
\label{order}
\end{equation}
These wave exponentials are solutions of a $\kappa$-Minkowski
wave equation~\cite{gacmaj}.
While wave exponentials of the
type $e^{i p^\mu x_\mu}$ would not combine in a simple way
(as a result of the $\kappa$-Minkowski noncommutativity relations),
for the ordered exponential one finds
\begin{equation}
(:e^{i p^\mu x_\mu}:) (:e^{i k^\nu x_\nu}:) =
:e^{i (p \dot{+} k)^\mu x_\mu}:
\quad.
\label{expprodlie}
\end{equation}
The notation ``$\dot{+}$" here introduced reflects the
behaviour of the mentioned ``coproduct"
composition of momenta in $\kappa$-Minkowski
spacetime:
\begin{equation}
p_\mu \dot{+} k_\mu \equiv \delta_{\mu,0}(p_0+k_0) + (1-\delta_{\mu,0})
(p_\mu +e^{\lambda p_0} k_\mu) ~. \label{coprod}
\end{equation}

As argued in Refs.~\cite{gacdsr} the nonlinearity
of the law of composition
of momenta should require an absolute
(observer-independent) momentum scale,
just like upon introducing a nonlinear law
of composition of velocities
one must introduce the absolute observer-independent scale of
velocity $c$. The inverse of the noncommutativity scale $\lambda$
should play the role of this absolute momentum scale.
This invites one to consider
the possibility~\cite{gacdsr}
that the transformation laws for energy-momentum
between different observers would have two invariants, $c$ and $\lambda$,
as required for a doubly-special-relativity
(deformed-Lorentz-symmetry) framework~\cite{gacdsr}.

In summary there is growing evidence
that in the rather general class of noncommutative
spacetimes described by Eq.(\ref{alllp}) it is never possible
to preserve Lorentz symmetry unmodified: in most cases Lorentz
symmetry is broken and in a few (possibly only in $\kappa$-Minkowski)
Lorentz symmetry is deformed.

\section{Spacetime foam and distance fluctuations}
I have so far focused on the fate of Lorentz symmetry in
quantum spacetime. Space experiments that can look for
effects possibly due to Planck-scale departures
from ordinary Lorentz symmetry will be discussed in the
next Section. In preparation for the discussion
of space experiments, I must first
here introduce another class of effects which has been
discussed in the quantum-gravity literature and for which
space experiments represent an important opportunity.
This originates from the
prediction of nearly all approaches to the unification
of general relativity and quantum mechanics that
at very short distances the sharp
classical concept of space-time should give way
to a somewhat ``fuzzy'' (or ``foamy'')
picture. In particular any given length/distance would be
affected by an irreducible level of uncertainty/fluctuations.

One way to characterize this operatively makes direct reference
to intereferometers~\cite{gacgwi,nggwi}.
Interferometers are the best tools for monitoring the distance
between test masses, and an operative definition
of quantum-gravity-induced
distance fluctuations can be expressed directly in terms
of strain noise in interferometers\footnote{Since
modern interferometers were planned to look for classical
gravity waves (gravity waves are their sought ``signal"),
it is reasonable to denominate as ``noise"
all test-mass-distance fluctuations
that are not due to gravity waves.}.
In achieving their remarkable accuracy modern interferometers
must deal with several classical-physics strain noise
sources ({\it e.g.}, thermal and seismic effects induce
fluctuations in the relative positions of the test masses).
Importantly, strain noise sources
associated with effects of ordinary quantum mechanics
are also significant for modern interferometers:
the combined minimization
of {\it photon shot noise} and {\it radiation pressure noise}
leads to a noise source which
originates from ordinary quantum mechanics~\cite{saulson}.
An operative definition of fuzzy distance can
characterize the corresponding quantum-gravity effects
as an additional source of strain noise.
A theory in which the concept of distance is
fundamentally fuzzy in this
operative sense would be such that
the read-out of an interferometer would still
be noisy (because of quantum-gravity effects)
even in the idealized limit in which all
classical-physics and ordinary-quantum-mechanics
noise sources are completely eliminated.
Just like the quantum properties of
the non-gravitational degrees of freedom of the apparatus
induce noise ({\it e.g.} the mentioned combination
of {\it photon shot noise} and {\it radiation pressure noise})
it is of course plausible that noise be induced
by the quantum properties of
the gravitational degrees of freedom of the apparatus
({\it e.g.} the distances between the test masses).

This operative definition of quantum-gravity-induced
distance fuzzyness immediately confronts us with a potentially
serious challenge, which is the central challenge of
all quantum-gravity-phenomenology research lines:
if indeed this distance fuzzyness is proportional
to (some power of) the Planck length $L_p$,
the smallness of $L_p \equiv 1/E_p \sim 10^{-35} m$
will automatically lead to very small effects. However,
modern interferometers
have a truly remarkable sensitivity to distance fluctuations
and it is not inconceivable that this
sensitivity would be sufficient for the detection
of fluctuations occurring genuinely at the Planck scale.
In order to support this observation with a simple
intuitive argument let us consider the possibility that the
distances $L$ betweeen the test masses of an interferometer be
affected by Planck-length
fluctuations of random-walk type
occurring at a rate of one per Planck time
($t_p = L_p/c \sim 10^{-44}s$).
It is easy to show~\cite{gacgwi}
that such fluctuations would induce strain noise
with power spectrum given by $L_p c L^{-2} f^{-2}$.
For $f \sim 100 Hz$ and $L\sim 1 Km$
(estimates relevant for some modern interferometers)
this corresponds to strain noise at the level $10^{-37} H\!z^{-1}$,
which is of course rather small because of the $L_p$ suppression
but is still well within the reach of the sensitivity of
modern interferometers.

Fluctuations genuinely at the Planck scale
(the simple scheme I used to illustrate my point
involves Planck-length fluctuations occurring
at a rate of one per Planck time)
can lead to an effect that, while being very small in absolute
terms, compares well with the sensitivity of modern interferometers.
This originates from the fact that random-walk fluctuations
do not fully average out. They have zero mean (in this sense
they do average out) but the associated standard deviation
grows with the time of observation (with the
random-walk-characteristic $\sqrt{t}$ dependence which
translates~\cite{gacgwi,rwold}
into the $f^{-2}$ dependence of the power spectrum).
A reasonable scale to characterize the time of observation
in interferometry is provided by $f^{-1}$
which, for $f \sim 100 H\!z$, is huge in Planck-time units.

\section{Space experiments and quantum properties of spacetime}
\subsection{Preliminaries on a simple model for Planck-scale
departures from Lorentz symmetry}
Studies of the fate of Lorentz symmetry in quantum gravity provide
a good prototype of quantum-gravity-phenomenology research line.
As discussed in the previous Sections, in several (though, of course,
not all) approaches to the quantum-gravity problem one finds some evidence
of departures from ordinary Lorentz symmetry.
Like other effects discussed in the quantum-gravity literature, the
ones associated with departures from Lorentz symmetry are very striking
from a conceptual perspective. While different intuitions for the quantum
gravity problem may lead to favouring one or another of these
effects, there is a general consensus that some strikingly
new effects should be present in quantum gravity.
It was however traditionally believed that even such
strikingly new effects (certainly leading to very characteristic signatures)
could not be tested because of their small magnitude, set by the small
ratio between the energy of the particles involved and the Planck energy
scale.
Work on quantum-gravity phenomenology has proven that
this old expectation is incorrect. And this point is very clearly
illustrated in the context of tests of Planck-scale departures
from Lorentz symmetry.

Rather than providing a more general discussion,
for simplicity I focus here on the possible emergence of
Planck-scale-modified dispersion relations,
\begin{equation}
E^2= m^2 + \vec{p}^2 + f(\vec{p}^2,E,m;L_p)
~,
\label{eq:disp}
\end{equation}
which are found in the large majority of quantum-gravity-motivated
schemes for deviations from ordinary Lorentz invariance
(see, {\it e.g.},
Refs.~\cite{grbgac,gampul,thiemLS,susskind,kpoinap,gacmaj,leeDispRel}).

If the function $f$ is nontrivial\footnote{For example, it would
be pointless to introduce an $f=L_p^2 [E^2 - \vec{p}^2 - m^2]^2$,
since then the dispersion relation (\ref{eq:disp}) would
be equivalent to $E^2= m^2 + \vec{p}^2$.} and
the energy-momentum transformation rules are unmodified (the familiar
Lorentz transformations) then clearly $f$ cannot have the exact
same structure for all inertial observers. In this case
Lorentz symmetry is necessarily ``broken",
in the sense clarified earlier.
In that case it is then legitimate to assume that,
in spite of the deformation
of the dispersion relation,
the rules for energy-momentum conservation would be undeformed.

If instead $f$ does have the exact
same structure for all inertial observers, then necessarily
the laws of transformation between observers must be deformed
(they cannot be the ordinary Lorentz transformation rules).
In this case Lorentz symmetry must be
deformed, in the sense of the doubly special relativity~\cite{gacdsr}
discussed earlier. There is no preferred frame.
The deformation of the laws of transformation between
observers impose that one must also necessarily~\cite{gacdsr} deform
the rules for energy-momentum conservation.

While the case of deformed Lorentz symmetry might exercise a stronger
conceptual appeal (since it does not rely on a preferred class of inertial
observers), for the purposes of this paper it is sufficient to consider
the technically simpler context of broken Lorentz symmetry.
Upon admitting a broken Lorentz symmetry
it becomes legitimate, for example, to adopt
a dispersion relation with leading-order-in-$L_p$ form
\begin{equation}
E^2 \simeq \vec{p}^2 + m^2 - \eta (L_p E)^n \vec{p}^2
~,
\label{eq:displead}
\end{equation}
without modifying the rules for energy-momentum conservation.
In (\ref{eq:displead}) $\eta$ is a phenomenological parameter
of order $1$ (and actually, for simplicity,
I will often implicitly take $\eta = 1$). $n$, the
lowest power of $L_p$ that leads to a nonvanishing
contribution, is model dependent.
In any given noncommutative geometry
one finds a definite value of $n$, and it appears to be equally
easy~\cite{gacdsr,dsrnext,grf03ess}
to construct noncommutative geometries with $n=1$ or with $n=2$.
In Loop Quantum Gravity one might typically
expect~\cite{grf03ess}
to find $n=2$, but certain scenarios~\cite{gampul}
have been shown\footnote{Note however that the Loop-Quantum-Gravity
scenario of Ref.~\cite{gampul} does not exactly lead to the dispersion
relation (\ref{eq:displead}): for photons ($m=0$) Ref.~\cite{gampul}
describes a polarization-dependent effect (birefringence).}
to lead to $n=1$.

I will use this popular Lorentz-symmetry breaking scenario,
with dispersion relation (\ref{eq:displead}) and unmodified
rules for energy-momentum conservation,
to illustrate how a tiny (Planck-length suppressed) effect
can be observed in certain experimental contexts.
The difference between the case $n \! = \! 1$ and
the case $n \! = \! 2$ is very significant from
a phenomenology perspective. Already with $n \! = \! 1$,
which corresponds to effects that are linearly suppressed by the Planck
length, the correction term in Eq.~(\ref{eq:displead}) is very small:
assuming $\eta \! \simeq \! 1$,
for particles with energy  $E \sim 10^{12} eV$
(some of the highest-energy particles we produce in laboratory)
it represents a correction of one part in $10^{16}$.
Of course, the case $n \! = \! 2$
pays the even higher price of quadratic suppression by the Planck length
and for $E \! \sim \! 10^{12} eV$ its effects are at the  $10^{-32}$
level.

\subsection{Gamma-ray bursts and Planck-scale-induced in-vacuo dispersion}
A deformation term of order $L_p^n E^n p^2$ in
the dispersion relation, such as the one in (\ref{eq:displead}),
leads to a small energy dependence of the speed of photons
of order $L_p^n E^n$, by applying the relation $v = dE/dp$.
An energy dependence of the speed of photons
of order $L_p^n E^n$ is completely negligible (both for $n=1$ and
for $n =2$) in nearly all physical
contexts, but, at least for $n=1$,
it can be significant~\cite{grbgac,billetal}
in the analysis of short-duration gamma-ray bursts that reach
us from cosmological distances.
For a gamma-ray burst a typical estimate of the time travelled
before reaching our Earth detectors is $T \sim 10^{17} s$.
Microbursts within a burst can have very short duration,
as short as $10^{-4} s$.
There is therefore one of the ``amplifiers" mentioned in Section~3:
the ratio between time travelled by the signal and time structure
in the signal is a (conventional-physics) dimensionless
quantity of order $\sim 10^{17}/10^{-4} = 10^{21}$.
It turns out that this ``amplifier" is sufficient to study
energy dependence of the speed of photons
of order $L_p E$ ($n=1$). In fact, some of the photons in these bursts
have energies in the $100 MeV$ range and higher.
For two photons with energy difference of order $100 MeV$ an $L_p E$
speed difference over a time of travel of $10^{17} s$
leads to a relative time-delay on arrival that
is of order $\Delta t \sim \eta T L_p \Delta E
 \sim 10^{-3} s$ (where $\Delta E$ is the difference between
 the energies of the two photons).
Therefore such a quantum-gravity-induced time-of-arrival delay
could be detected~\cite{grbgac,billetal}
upon comparison of the structure of the signal
in different energy channels.

A space telescope, the GLAST~\cite{glast}
gamma-ray satellite telescope (scheduled to start taking data in 2006),
is presently considered to be the best opportunity to
look for this effect.
AMS~\cite{glast}, on the International Space Station, could also
contribute to these studies.

While GLAST (and possibly AMS) should have sufficient sensitivity
to look for the $n=1$ case, {\it i.e.} $L_p E$ corrections,
the much weaker energy dependence of the speed of light
found in the case $n=2$  ($L_p^2 E^2$ corrections)
is clearly beyond the reach of GLAST.
For $n=2$ the
time-of-arrival-difference (again in the illustrative example
of two photons with energies in the $100 MeV$ range)
comes out to be of order $10^{-18} s$,
which is not only beyond the sensitivities achievable with GLAST
but also appears to be unreachable
for all foreseeable gamma-ray observatories.

Some access to effects characterized by the $n=2$ case
could be gained by exploiting the fact that, according to
current models~\cite{grbNEUTRINOnew},
gamma-ray bursters should also emit a substantial amount of
high-energy neutrinos.
Models of gamma-ray bursters predict in particular a substantial
flux of neutrinos with energies of about $10^{14}$ or $10^{15}$ $eV$.
Comparing, for example, the times of arrival of these neutrinos emitted
by gamma-ray bursters to the corresponding times of arrival of
low-energy photons, the case $n=1$ would predict a
huge time-of-arrival difference ($\Delta t \sim 1 year$)
and even for the case $n=2$ the time-of-arrival
difference could be significant ({\it e.g.} $\Delta t \sim 10^{-6} s $).

Current models of gamma-ray bursters also predict some
production of neutrinos
with energies extending to the $10^{19} eV$ level.
For such ultra-energetic neutrinos one would expect
an even more significant signal, possibly at the
level $\Delta t \! \sim \! 1 s $ for $n \! = \! 2$.
A $1 s$ timing accuracy will surely be
comfortably\footnote{For this strategy relying on ultra-high-energy
neutrinos the delicate point is clearly not timing, but rather the
statistics (sufficient number of observed neutrinos)
needed to establish a robust experimental result.
Moreover, it appears necessary to understand gamma-ray bursters
well enough to establish whether there are typical
at-the-source time delays.
For example, if the analysis is based
on a time-of-arrival comparison between the first (triggering)
photons detected from the burster and the first neutrinos
detected from the burster it is necessary to establish that
there is no significant at-the-source effect such that
the relevant neutrinos and the relevant photons are emitted
at significantly different times.
The fact that this ``time history"
of the gamma-ray burst must be obtained only with precision
of, say, $1 s $ (which is a comfortably large time scale with respect
to the short time scales present in most gamma-ray bursts)
suggests that this understanding should be achievable in the
not-so-distant future.} available to the EUSO~\cite{euso}
observatory, being planned for the International Space Station.
EUSO is expected to be able to observe high-energy
neutrinos (even with energies higher than $10^{19}$ $eV$)
and one could, for example, attempt to correlate such detections of
high-energy neutrinos with corresponding detections
of lower-energy gamma-ray-burst particles ({\it e.g.} $MeV$ photons).

\subsection{Cosmic rays and Planck-scale-modified thresholds}
Let us now consider another significant prediction
that follows from the dispersion relation (\ref{eq:displead}).
While in-vacuo dispersion, discussed in the preceding Subsection,
only depends on the deformation of the dispersion relation\footnote{The
dispersion relation (\ref{eq:displead}) can also be implemented in
a doubly special relativity (deformed Lorentz symmetry)
scenario~\cite{gacdsr}. The in-vacuo-dispersion analysis discussed
in the preceding Subsection applies both to Lorentz-symmetry-breaking
and Lorentz-symmetry-deformation scenarios adopting (\ref{eq:displead}).
When (\ref{eq:displead}) is adopted in a Lorentz-symmetry deformation
scenario it is necessary~\cite{gacdsr} to consistently modify the laws
of energy-momentum conservation. Therefore the analysis of
Planck-scale-modified thresholds discussed in this Subsection,
which assumes unmodified laws of energy-momentum conservation,
does not apply to the scenario in which (\ref{eq:displead})
is adopted in a Lorentz-symmetry deformation scenario.
Planck-scale-modified thresholds are present also in the
case of  Lorentz-symmetry deformation, but there are significant
quantitative differences~\cite{gacdsr}.},
the effects considered in this Subsection, the so-called ``threshold
anomalies"~\cite{gactp,nguhecr},
also depend
on the rules for energy-momentum conservation, which are not modified
in the Lorentz-symmetry-breaking scenario I am considering.

Certain types of energy
thresholds for particle-production processes may be sensitive
to the tiny $L_p^n E^n p^2$ modification of
the dispersion relation I am considering.
Of particular interest is the analysis of the photopion
production process $p + \gamma \rightarrow p + \pi$ when the
incoming proton has high energy $E$ while the incoming
photon has much smaller energy $\epsilon$ ($\epsilon \ll E$).
In fact, adopting the modified dispersion relation (\ref{eq:displead})
and imposing ordinary (unmodified) energy-momentum conservation
one finds~\cite{gactp} that, for fixed photon energy $\epsilon$,
the process is allowed when $E > E_{th}$, with $E_{th}$
given by the modified threshold relation
\begin{equation}
E_{th} \simeq {(m_p + m_\pi)^2 - m_p^2 \over 4 \epsilon}
- \eta {L_p^n E_{th}^{2+n} \over 4 \epsilon } \left(
{m_p^{1+n} + m_\pi^{1+n} \over (m_p + m_\pi)^{1+n}} -1 \right)
~.
\label{deltaeth}
\end{equation}
Of course, the ordinary threshold relation for photopion
production is obtained by taking the $L_p \rightarrow 0$ limit
of (\ref{deltaeth}), in which the correction of
order $L_p^n E_{th}^{2+n}/\epsilon$ disappears. The correction of
order $L_p^n E_{th}^{2+n}/\epsilon$ can be relevant
for the analysis of ultra-high-energy cosmic rays.
A characteristic feature of the expected cosmic-ray spectrum,
the so-called ``GZK limit", depends on the evaluation of the
minimum energy required of a cosmic ray in order to produce pions
in collisions with cosmic-microwave-background
photons.
According to ordinary Lorentz symmetry this threshold energy
is around $E_{th}  \! \simeq  \! 5 {\cdot} 10^{19} eV$,
and cosmic rays emitted with energy in
excess of this value should loose the excess
energy through pion production before reaching the Earth.
Strong interest was generated by the
observation~\cite{kifu,ita,aus,gactp,jaco,alfaro}
that the Planck-scale-modified threshold relation (\ref{deltaeth})
leads, for positive $\eta$, to a higher estimate of the
threshold energy, an upward shift of the GZK limit.
This would provide a description of the observations of the high-energy
cosmic-ray spectrum reported by AGASA~\cite{agasa},
which can be interpreted as an indication of
a sizeable upward shift of the GZK limit.
Both for the case $n= 1$ and  for the case $n= 2$
the Planck-scale-induced upward shift would be large
enough~\cite{kifu,ita,aus,gactp,jaco,grf03ess,alfaro}
for quantitative agreement with the cosmic-ray observations
reported by AGASA.

There are other plausible explanations for
the AGASA ``cosmic-ray puzzle",
and the experimental
side must be further explored, since another
cosmic-ray observatory, HIRES, has not confirmed the AGASA results.
The situation will become clearer with planned more powerful
cosmic-ray observatories,
such as the Pierre Auger (ground) Observatory,
which will soon start taking data,
and EUSO (which, as mentioned, is planned for the International
Space Station).

\subsection{Interferometry
and Planck-scale-induced in-vacuo dispersion}
Planck-scale
modifications of Lorentz symmetry
could also  affect certain
laser-interferometric setups, as
L{\"{a}}mmerzahl and I recently
observed~\cite{gaclaem}.
Our observation is based on the idea of operating
a laser-light interferometer
with two different frequencies\footnote{Modern interferometers
achieve remarkable
accuracies also thanks to an optimization of all experimental
devices for response to light of a single frequency.
The requirement of operating with light at two different
frequencies is certainly a challenge for the realization
of interferometric setups of
the type proposed in Ref.~\cite{gaclaem}.
This and other practical concerns are not discussed here.
The interested reader can find a preliminary discussion
of these challenges in Ref.~\cite{gaclaem}.},
possibly obtained from a single laser
beam by use of a ``frequency doubler''
(see {\it e.g.} \cite{Sauter96}).

A description of some interferometric setups that could
be used for this purpose is given in
Ref.~\cite{gaclaem}.
At least at a simple-minded level
of comparison between
the magnitude of the Planck-scale-induced phase difference
and the phase-difference sensitivity of LIGO/VIRGO-type
or LISA-type interferometers
we find some encouragement for this proposal.

In addition to studies of Planck-scale
modifications of Lorentz symmetry, laser-light interferometry
could also be used to explore the possibility
of quantum-gravity-induced distance fluctuations,
in the sense here discussed in Section~7.
The strategy that should be followed is rather clear since
in Section~7 I characterized physically the distance fluctuations
directly in terms of interferometry.
One should look for excess strain noise, noise in excess with
respect to the one expected from classical mechanics and
ordinary quantum mechanics. Whereas for
Planck-scale departures from Lorentz
symmetry there is a natural phenomenology in terms of
modified dispersion relations, and one manages to find
evidence in support of such dispersion relations
in certain quantum-gravity proposals (at least
in noncommutative spacetimes and in Loop Quantum Gravity),
it is much more difficult to develop a phenomenological
model for distance fluctuations. The derivation of the much
discussed ``uncertainty principle for lengths/distances''
in quantum gravity theories remains at a rather heuristic level,
and it remains to be established which (if any) relation
connects the new uncertainty principle to the structure of
distance fluctuations.
The simple-minded description of random-walk distance fluctuations
given in Section~7 does not necessarily provide a reliable
estimate and a reliable formula for quantum-gravity-induced
interferometric noise~\cite{polonpap,gacgwi},
but it shows
that effects genuinely at the Planck scale
(Planck-length fluctuations with Planck-time frequency)
could lead to observably large noise levels.
Moreover, the random-walk distance-fluctuation picture also provides
insight on another perhaps unexpected feature of
quantum-gravity-induced noise: this noise might be
peaking at low frequencies~\cite{polonpap,gacgwi},
rather than at the characteristic
quantum-gravity frequency scale (the Planck frequency, given
by the inverse of the Planck time).
Through the example of random-walk fluctuations one can
see that the value of the frequency at which noise
is highest is not simply set by the inverse of the characteristic
time scale of the fluctuations: random-walk fluctuations
will always lead to noise peaking at low frequencies,
independently of the time scale of fluctuations.
[The fact that it peaks in the low-frequency limit is a direct
consequence~\cite{rwold,gacgwi}
of the fact that the standard deviation of random-walk
fluctuations grows in time, with the
characteristic $\sqrt{t}$ behaviour.]

As long as we lack a satisfactory model of distance fluctuations
we cannot do any better than look for excess noise
at low frequencies. In this respect space interferometers
can be important, since in space some of the low-frequency
noise sources present in on-ground interferometers are
absent. In particular, the LISA space interferometer
is expected to achieve a remarkably low level of
low-frequency noise, and one could perhaps hope that
this might open a window of opportunity for the
discovery of quantum-gravity-induced low-frequency
excess noise.
Moreover, in the development of this quantum-gravity-phenomenology
research lines based on laser-light interferometry
an important role could be played
by the operation of other lasers and
cavities in space laboratories~\cite{laemopti,buchlaser},
such as the International Space Station.
The (relatively) quiet environment
of space could be ideally suited for the development
of new laser-interferometric techniques,
such as the ones needed for the proposal of
Ref.~\cite{gaclaem}.


\baselineskip 12pt plus .5pt minus .5pt

\vfil

\end{document}